**Attenuation of acoustic waves in glacial ice and salt domes**

P. B. Price

Physics Department, University of California, Berkeley, CA 94720, USA

Two classes of natural solid media – glacial ice and salt domes – are under consideration as media in which to deploy instruments for detection of neutrinos with energy $\geq 10^{18}$ eV. Though insensitive to $10^{11}$ to $10^{16}$ eV neutrinos for which observatories (e.g., AMANDA and IceCube) that utilize optical Cherenkov radiation detectors are designed, radio and acoustic methods are suited for searches for the very low fluxes of neutrinos with energies $>10^{17}$ eV. This is because, due to the very long attenuation lengths of radio and acoustic waves in ice and salt, detection modules can be spaced very far apart. In this paper, I calculate the absorption and scattering coefficients as a function of frequency and grain size for acoustic waves in glacial ice and salt domes and show that experimental measurements on laboratory samples and in glacial ice and salt domes are consistent with theory. For South Pole ice with grain size ~0.2 cm at -51°C, scattering lengths are calculated to be 2000 km and 25 km at 10 kHz and 30 kHz, respectively, and the absorption length is calculated to be 9 ± 3 km at frequencies above ~100 Hz. For NaCl (rock salt) with grain size 0.75 cm, scattering lengths are calculated to be 120 km and 1.4 km at 10 kHz and 30 kHz, and absorption lengths are calculated to be 3 x $10^4$ km and 3300 km at 10 kHz and 30 kHz. Existing measurements are consistent with theory. For ice, absorption is the limiting factor; for salt, scattering is the limiting factor.

## 1. Introduction

IceCube, the first kilometer-scale neutrino observatory, is now under construction in South Pole ice. It will have 80 strings, each instrumented with 60 phototubes at depths from 1400 to 2400 m in a 1 km x 1 km x 1 km volume [*Ahrens et al.*, 2004]. The timing and strength of flashes of optical Cherenkov radiation recorded in the phototubes enable the direction and energy of an interacting neutrino to be reconstructed. Because the neutrino flux falls off so rapidly with energy, an observatory far larger than 1 km$^3$ will be required in order to detect neutrinos with energies above ~$10^{18}$ eV. But those ultrahigh energy neutrinos are predicted to have great cosmological significance and may reveal new physics beyond the standard model. The GZK (Greisen-Zatsepin-Kuzmin) neutrinos, with energies ~$10^{17}$ – $10^{20}$ eV, which should result from interactions of high-energy cosmic ray protons with the 2.7°K cosmic background photons, are an example. Estimates indicate that no more than ~1 GZK neutrino per year can be detected with IceCube. With a future observatory of effective volume more than an order of magnitude larger than IceCube, it may be possible to probe new physics and predictions of some cosmological models. Simulations of detection rates of such neutrinos have led to a straw design of a hybrid optical/radio/acoustical neutrino detector with $>10^2$ km$^3$ effective volume [*Vandenbroucke et al.*, 2005a]. At its core would be the IceCube optical array; extending laterally outward would be a combination of radio receivers and acoustic transducers, which could be sparsely spaced because the attenuation lengths of radio and acoustic waves from ultrahigh-energy neutrino interactions are expected to be much greater than for visible light. The occasional detection of an ultrahigh-energy neutrino with optical modules in coincidence with radio and acoustical modules would validate the radio and acoustical techniques, which are insensitive to neutrino interactions except at energies above ~$10^{17}$ eV.

With the idea of a future gigantic neutrino observatory in mind, some years ago I discussed the absorption and scattering of acoustic waves in the 2.8-km-deep glacial ice at the South Pole [*Price,* 1993]. Since then, a number of workers have been investigating the feasibility of detecting ultrahigh-energy neutrino interactions in underground NaCl domes, of typical dimensions ~3 km x ~3 km x ~3 km, using either radio or acoustic receivers. In support of these efforts, in this paper I discuss theory and measurements of acoustic attenuation in laboratory samples and *in-situ* in natural media for both ice and salt. I consider only longitudinal (pressure) waves, since the efficiency for production of transverse (shear) waves associated with an electromagnetic cascade is low and uncertain.

## 2. Absorptivity of optical Cherenkov radiation at 200 to 2000 nm in ice and salt

Before discussing acoustic propagation, a brief remark on propagation of optical Cherenkov radiation is in order. High-purity Antarctic ice [*Askebjer et al.*, 1997] and high-purity NaCl [*Palik*, 1985] happen to be the two most optically transparent solids known. Figure 1 compares laboratory-grown ice, ice at various depths at the South Pole, and laboratory-grown salt in the UV, visible, and IR. Taking into account that the intensity of Cherenkov light goes as $\lambda^{-2}$ ($\lambda$ is wavelength), it is clear that, despite the great transparency of NaCl in the infrared, more light from Cherenkov radiation by a neutrino-induced cascade or muon would be detected in a phototube immersed in Antarctic ice than in a salt dome. The drawback to using visible light to detect ultrahigh-energy neutrino interactions in either ice or salt is that absorption lengths are typically only ~$10^2$ m. This limits the size of an observatory on the grounds of cost, since the detection and reconstruction of events requires that the spacing of phototubes be ≤ absorption length. To fill $10^2$ km$^3$ of ice or salt with phototubes would thus be prohibitively expensive.

## 3. Conversion of ionization energy into acoustic energy

The conversion of ionization energy into acoustic energy is orders of magnitude less efficient than its conversion into visible light. The thermoacoustic model [*Askariyan et al.*, 1979] is based on the assumption that the energy in a neutrino-induced electromagnetic cascade in a small volume of liquid or solid will overheat that volume, leading to a pressure pulse. The amplitude of this pulse is a measure of the cascade energy. The pulse propagates radially from the axis like a radially expanding flat disk, providing a measure of the direction of the incoming cascade. The frequency of the pressure wave is in the 10 to 60 kHz range and is related to the transverse dimension of the cascade. Table 1 compares the figures of merit for seawater, ice, and NaCl as media for conversion of ionization energy in a cascade into the energy in the resulting pressure wave. Ice is a factor 7.3 more effective than seawater (and a factor 10 more effective than lake water), and NaCl is a factor 2.6 more effective than ice.

In addition to conversion efficiency, ambient noise is also an important consideration. The noise level in the ocean, due to wind, waves, and organisms, is much higher than in glacial ice or in a salt dome. See section 9. Finally, one has to consider available volume. As indicated by the last row in Table 1, the practical maximum volume of ocean that could be instrumented is larger than for ice, and the practical volume of ice is in turn larger than that for a salt dome.

## 4. Refraction and absorptivity of acoustic waves in the ocean

Scattering of acoustic waves in the ocean is negligible. Refraction depends on the gradient of acoustic velocity, which is a function of both temperature and density. To a good approximation, the refractive index and sound speed vary only in the vertical (*z*) direction. This leads to two equations, valid provided the refractive index does not change much over the distance of one wavelength. The first, Snell's law for refraction of an acoustic wave, relates the angle $\theta$ of a particular ray to the vertical to the initial angle $\theta_0$ when at a depth where the refractive index is *n*:

$$\frac{\cos\theta}{\cos\theta_0} = \frac{v}{v_0} \equiv \frac{1}{n} \qquad (1)$$

The second states that the curvature of a ray, $d\theta/ds$, is directly proportional to the velocity gradient:

$$\frac{d\theta}{ds} = -K\frac{dv}{dz} = -\frac{\cos\theta_0}{v_0}\frac{dv}{dz} = \frac{1}{R} \qquad (2)$$

where *R* is the radius of curvature. For a constant velocity gradient, the curvature along any given ray is constant, so the ray is the arc of a circle.

Figure 2 shows the frequency-dependent contributions to absorptivity in seawater. At frequencies above ~100 kHz, it behaves in the same manner as fresh water, with energy dissipation resulting from viscosity of the water. At lower frequencies, pressure-dependent chemical reactions can cause acoustic absorption. Pressure waves shift chemical equilibrium between dissolved molecular compounds and their dissociated ions, taking energy from the waves. For frequencies up to nearly 10 kHz, absorption is caused primarily by the presence of boric acid, whose dissociation has a time constant such that energy-absorbing shifts in its equilibrium cannot take place in times shorter than a few tenths of a millisecond. At higher frequencies, up to a few hundred kHz, a dissociation process associated with dissolved $MgSO_4$ produces the energy losses. The equations for $B(OH)_3$ and $MgSO_4$ are:

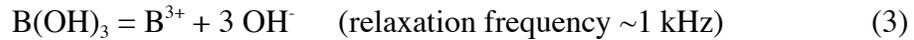

$$B(OH)_3 = B^{3+} + 3\ OH^- \quad \text{(relaxation frequency ~1 kHz)} \qquad (3)$$

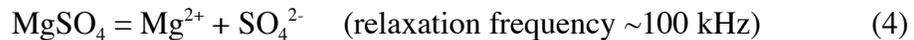

$$MgSO_4 = Mg^{2+} + SO_4^{2-} \quad \text{(relaxation frequency ~100 kHz)} \qquad (4)$$

## 5. Acoustic refraction and scattering in ice

### 5.1. Refraction in firn ice

Refraction of acoustic waves in firn ice is qualitatively similar to that in lunar soil in that both curve upward due to the monotonic increase of density with depth. Seismologists found that low-frequency propagation of seismic waves from meteorite impacts on the lunar surface propagated completely around the moon due to the waveguide effect of the density gradient in the firn.

If the velocity as a function of depth in firn has been measured, Eqs. (1) and (2) can be used to derive the ray paths. *Albert* [1998] has used the equations, together with wavenumber integration, to model seismic noise propagation at 3, 10, and 30 Hz as a function of firn depth and horizontal distance from the South Pole station. At those low frequencies he found that the reduction in wave amplitude with depth and distance is due mainly to wavefront spreading and to upward curvature in the firn, not to absorption or scattering. However, at frequencies relevant to detection of neutrino-induced cascades, we have neither experiment nor theory to guide in modeling attenuation until a depth below firn closeoff is reached. From there downward, the discussions in sections 5.2 and 5.3 on scattering and in section 6 on absorption apply.

5.2. Scattering in bubbly ice

The mean diameter of air bubbles decreases monotonically with depth from the bottom of the firn layer, at ~100 m, to ~1400 m, below which all bubbles have converted into the solid clathrate phase. The density of the clathrate crystals is close to that of the ice, so the scattering of acoustic waves by clathrates is negligibly small. In the bubbly regime, the mean spacing between bubbles is much greater than their linear dimensions, and they act as independent scatterers. For frequencies of interest here, the Rayleigh approximation ($\lambda/2\pi d_b > 1$, where $\lambda$ is wavelength and $d_b$ is bubble diameter) is valid. The scattering coefficient for a
longitudinal wave depends on bubble concentration, $n_0$, as well as on $d_b$ and frequency [*Price*, 1993].

$$b_{bubble} \text{ [m}^{-1}\text{]} = 2.68 \times 10^{-10} (n_o/200 \text{ cm}^{-3}) (d_b/0.02 \text{ cm})^6 (f/10 \text{ kHz})^4 \qquad (5)$$

5.3 Scattering at grain boundaries

In an elastically isotropic solid such as glass, the speed of elastic waves is independent of direction, and there is no scattering. In crystalline solids, scattering occurs at grain boundaries, across which the acoustic speed changes abruptly. At all depths in terrestrial glacial ice the crystal structure is hexagonal, and the acoustic properties are functions of the five elastic constants $c_{11}$, $c_{12}$, $c_{13}$, $c_{33}$, and $c_{44}$. There are three distinct scattering regimes, depending on the magnitude of the parameter $\lambda/2\pi d$, where $\lambda$ is acoustic wavelength and $d$ is the mean diameter of the grain in polycrystalline ice. In the Rayleigh regime ($\lambda/2\pi d > 1$), the scattering coefficient $b(d,f)$ is proportional to $d^3 f^4$. For longitudinal and transverse waves in a solid with hexagonal structure, *Merkulov* [1957a] derived the expressions

$$b_L = \frac{2\pi^4 d^3 f^4}{1350 \rho^2 v_L^3} \left( \frac{a_1}{v_L^5} + \frac{b_1}{v_T^5} \right) \qquad \text{and} \qquad b_T = \frac{2\pi^4 d^3 f^4}{1350 \rho^2 v_T^3} \left( \frac{a_2}{v_L^5} + \frac{b_2}{v_T^5} \right) \qquad (6)$$

where the dependences of the coefficients $a_1$, $b_1$, $a_2$, and $b_2$ are given by *Price* [1993]. The two terms in parentheses take into account conversion between the two modes during scattering at grain boundaries. In South Pole ice with $v_L$ = 3920 km/s and $v_T$ = 1995 km/s, in the Rayleigh regime,

$$b_L = 5 \times 10^{-4} (d/0.2 \text{ cm})^3 (f/10 \text{ kHz})^4 \text{ [km}^{-1}\text{]} \qquad (7)$$

In the stochastic regime ($0.5 < \lambda/2\pi d < 1$), the scattering deviates from the Rayleigh law, increasing only as $d \cdot f^2$ [*Merkulov*, 1957a]. The result for ice is

$$b_L = 6.2 (d/0.2 \text{ cm}) (f/500 \text{ kHz})^2 \text{ [m}^{-1}\text{]} \qquad (8)$$

This weaker dependence on $d$ and $f$ is caused by the coherent nature of the scattering process. For $d = 0.2$ cm, Eq. (8) holds in the frequency regime 310 kHz $< f <$ 620 kHz.

In the geometric regime ($\lambda/2\pi d < 0.5$), the quadratic dependence of $b_T$ on frequency disappears, and the scattering mechanism approaches a diffusive process [*Merkulov*, 1957b]. The wave velocity shows a discontinuity in magnitude and direction at each grain boundary, and some of the energy is reflected at the boundary. The resulting scattering is independent of frequency and inversely proportional to mean grain diameter:

$$b_L = \langle R \rangle /d \tag{9}$$

For the case of ice, for which the elastic anisotropy of individual crystal grains is small, the average reflection coefficient is given by $\langle R \rangle = 0.068$ (see *Price* [1993] for details), and

$$b_L \, [\text{m}^{-1}] = 6.8/d[\text{cm}] \tag{10}$$

For $d = 0.2$ cm, this leads to $b_L \approx 34$ m$^{-1}$, valid at frequencies above ~620 kHz. Note that the energy is not lost, merely redirected. For all cases of interest ($d <$ 2 cm), the frequency at which scattering reaches its limiting value is above 100 kHz.

Figure 4 shows the dependence of grain-boundary scattering on grain diameter and frequency for frequencies relevant to neutrino-induced hydrodynamic wave propagation. In the top 600 m, the grain diameter is probably no larger than ~0.2 cm, and it increases with depth. Since no ice core deeper than a few hundred m in South Pole ice has been taken, the coefficient in deeper ice is not known. One sees from the figure that acoustic scattering in ice with 0.2 cm grain diameter is extremely low: in the frequency regime where the acoustic energy of a neutrino-induced cascade is concentrated, the scattering length is 2000 km at 10 kHz, 25 km at 30 kHz, and 2 km at 60 kHz.

## 6. Acoustic absorption in ice

For longitudinal waves in very cold South Pole ice, where the temperature is -51°C near the surface and rises slowly to -46°C at 1000 m, the dominant relaxation mechanism is proton reorientation, whereas in warm ice near bedrock, where the temperature reaches -9°C [*Price et al.*, 2002], grain boundary sliding dominates. We consider only proton reorientation, which dominates in South Pole ice at depths less than ~2000 m where an acoustic ultrahigh-energy neutrino astrophysics observatory is under consideration for development after the completion of IceCube [*Vandenbroucke et al.*, 2005a]. The contribution of proton reorientation to absorption is independent of grain size, whereas the contribution of grain boundary sliding to absorption in warm ice does depend on grain size. The dipole moment of the H$_2$O molecule may assume one of six orientations. Passage of an elastic wave may result in movement of some of the protons, causing rotation of dipole moments into favored orientations. The rotation proceeds by the movement of orientational defects, whose concentration increases with temperature, from one hydrogen bond to another. As Fig. 5(a) shows, D and L defects denote bond sites occupied by two protons (D for the German word "doppel" or double) and no protons (L for "leer" or empty). In addition, ionization defects, OH$^-$ (only 1 proton attached to an oxygen) and H$_3$O$^+$ (3 protons attached to an oxygen) can migrate *via* a shift of a proton along the same hydrogen bond from one oxygen to another in response to an acoustic wave. Defect reorientation and migration lead to "internal friction", with a relaxation time $\tau_m(T) = \tau_o \exp(U/kT)$, where $U$ is the activation energy for mechanical relaxation. The logarithmic decrement for energy loss due to the acoustic wave given by

$$\delta = \delta_{max} \frac{4\pi f \tau_m}{1 + 4\pi^2 f^2 \tau_m^2} \qquad (11)$$

where $\delta_{max}$ is a constant, independent of frequency, with a value that depends on the wave mode and the direction of propagation. For polycrystalline ice with randomly oriented grains, an experimentally determined average value can be taken. Due to the low temperature in the top ~600 m of South Pole ice, there is essentially no shear at those depths, which validates the assumption of randomly oriented grains and of an average value of $\delta_{max}$. The absorptivity and absorption length are then given by

$$a_L \text{ [m}^{-1}\text{]} \equiv 1/\lambda_a = \delta f / v_L \qquad (12)$$

where $\delta$ has a temperature dependence contained in $\tau_m(T)$.

Table 2 summarizes the results of three sets of measurements of $U$, $\tau_o$, and $\delta_{max}$ obtained from internal friction measurements of laboratory ice samples. The last two columns give the values of absorption length at near-surface temperature (-51°C) and at 1000 m (-46°C) in South Pole ice, calculated from the measurements of $U$, $\tau_o$, and $\delta_{max}$.

Figure 6 shows curves of absorptivity as a function of frequency at several temperatures, calculated from data of *Kuroiwa* [1964]. As the table shows, his values of absorption length fall between those generated from the data of *Schiller* [1958] and of *Oguro et al.* [1982]. In addition, several field measurements have been reported. *Bentley and Kohnen* [1976] measured seismic attenuation of longitudinal waves in ice at depths 100 to 500 m and a temperature –28°C near Byrd Station at 136 Hz (solid circle in Fig. 6), and they used reflection shooting in Antarctica and Greenland to infer internal friction at 50 to 200 Hz. Their results were similar to those obtained by *Brockamp and Kohnen* [1965] at –22°C in Greenland ice (solid triangle). Both points agree well with the calculations for the respective temperatures.

One could make a case for adopting the data of *Oguro et al.* [1982] on the ground that they were obtained some 18 years after Kuroiwa's data. To be conservative, I use Kuroiwa's data and take the spread among the three authors as an indication that the uncertainty is ~±30%.

**7. Scattering of acoustic waves in NaCl**

Salt evaporite beds such as are used as repositories for nuclear waste have NaCl layers >100 m thick separated by other minerals. Impurity contents include >1% water, mostly in liquid inclusions, as well as horizontal thin beds of clay, silt, and anhydrite ($CaSO_4$), and nuggets of sulfur, calcite, and $BaSO_4$ – all of which make evaporite beds unattractive for acoustic detection of neutrino cascades. Salt domes are purer than evaporite beds and probably have much longer absorption lengths. Several mines are known to consist of >99% polycrystalline NaCl and contain only 2 to 40 ppm water. The salt grains dynamically recrystallize while the dome is plastically deforming due to its lower density than that of surrounding rock. Scattering occurs at boundaries of recrystallized grains with random orientation. Boundaries of subgrains inside grains have misorientations < 1° where scattering is negligible. Scattering can also occur at interfaces between NaCl and layers of clay or anhydrite.

As Fig. 5(b) shows, the crystal structure of NaCl is cubic. The equations for grain-boundary scattering for a crystal with cubic structure are slightly different from those for the hexagonal structure [*Merkulov*, 1957a,1957b]. Eqs. (13) and (14) apply in the Rayleigh and stochastic regime, respectively:

$$b_L(\text{Rayleigh}) = \frac{4\pi^4 d^3 f^4}{1125} \frac{(c_{11} - c_{12} - 2c_{44})^2}{\rho^2 v_L^3} \left[\frac{2}{v_L^5} + \frac{3}{v_T^5}\right] = 2 \times 10^{-5} \cdot \left(\frac{d}{1\,\text{cm}}\right)^3 \left(\frac{f}{10^4}\right)^4 m^{-1} \quad (13)$$

$$b_L(\text{stochastic}) = \frac{16\pi^2 df^2}{525} \frac{(c_{11} - c_{12} - 2c_{44})^2}{\rho^2 v_L^6} = 3.6 \times 10^{-4} \cdot \left(\frac{d}{1\,\text{cm}}\right)\left(\frac{f}{10^4}\right)^2 m^{-1} \quad (14)$$

Figure 7 gives my calculated values of scattering coefficient in polycrystalline NaCl as a function of frequency for mean grain diameters 0.2 to 2 cm. The points labeled with triangles show my analyses of acoustic amplitudes at frequencies 20 to 100 kHz at distances 7 and 71 m from a transmitter at a depth of 500 m in a salt repository in Morsleben, Germany [*Manthei and Eisenblätter*, 2005]. Although grain size was not measured, I infer from the similarity of the scattering coefficients to the dashed curve that the mean grain diameter was ~1 cm. Points labeled with solid deltas were measured at 58 to 64 kHz at a depth of 470 m in the Hockley Salt Mine, Texas. The scattering coefficient is a factor ~10 higher than would be expected for the measured mean grain size of 0.6 to 0.8 cm. Manfred Fink (priv. comm.) pointed out to me that a 0.1 to 1 mm-thick layer of powdery anhydrite is present in all grain boundaries. In my opinion, this could provide enough additional scattering to account for the high values.

**8. Absorption of acoustic waves in NaCl**

The absorptive mechanism that dominates in ice does not apply in NaCl, because the structure of NaCl has no protons and no hydrogen bonds. Experimental studies have shown that a potentially important relaxation mechanism, in which dislocations are driven by an acoustic wave, can also be neglected if the NaCl crystals have been well annealed. In salt repositories, the strain rate is so slow that ongoing annealing at ambient temperature maintains a fairly low dislocation density. The dominant absorption mechanism is due to phonon-phonon interactions, which were first studied by *Akhieser* [1939] and quantitatively evaluated by *Merkulov* [1970], *Sahasrabudhe and Lambade* [1998], and others. In this process, through anharmonic interactions, the entire thermal phonon gas extracts energy from the phonons that compose the acoustic wave. Sahasrabudhe and Lambade compared their calculations with data for NaCl, NaF, and LiF in three crystallographic directions at temperatures from 80 K to 300 K. The absorptivity due to the Akhieser relaxation mechanism is

$$a_L[m^{-1}] = \frac{E_0 D}{6\rho v_L^3} \frac{4\pi^2 f^2 \tau_L}{1 + 4\pi^2 \tau_L^2} \quad (15)$$

which becomes, for $2\pi f \tau_L \ll 1$,

$$a_L = 2E_0 D\pi^2 f^2 \tau / 3\rho v_L^3 \qquad (16)$$

where $E_0$ is the thermal energy density, $\rho$ is the density, $v_L$ is the longitudinal wave velocity, and $D$ is the nonlinearity constant, which is derived from the second- and third-order elastic constants. The relaxation time $\tau_L$ for exchange of acoustic and thermal energies is given by

$$\tau_L = 6K/C_v \langle V \rangle^2 \qquad (17)$$

where $K$ is the thermal conductivity (carried entirely by phonons if the crystal is nonconducting), $C_v$ is the specific heat per unit volume, and $\langle V \rangle$ is the average Debye velocity. For details, see *Kor and Mishra* [1975].

Using the $f^2$ dependence in Eq. (16), I calculated the dotted line in Fig. 7 labeled absorption (phonon-phonon) by scaling laboratory data for acoustic absorptivity of NaCl [*Sahasrabudhe and Lambade*, 1998] from the megahertz to the kilohertz regime. The absorptivity has been shown experimentally to be independent of temperature to within ±30% for $T > 60$ K.

## 9. Ambient acoustic noise in glacial ice and salt domes

At the highest frequency, ~30 Hz, the seismic station 200 m deep in ice ~8 km from South Pole has the lowest noise of all stations in the Global Seismic Network [*R. Butler*, priv. comm.] Glacial ice at the South Pole flows toward the ocean at a surface rate of nearly 10 m/yr, whereas at bedrock the rate is close to zero. Most of the shear occurs near bedrock where the temperature is warm enough for plastic deformation to occur. If the shear is of the stick-slip type, discrete acoustic emissions may occur close to or at bedrock. Laboratory experiments on creep deformation of ice at -10 C show that acoustic emissions are broadly distributed in frequency from ~0.01 to 1 MHz [*Weiss and Grasso*, 1997]. Noise as a function of frequency and time will be measured in South Pole ice with a small array in 2006 [*Vandenbroucke et al.*, 2005b].

Creep due to slow upthrust of a salt dome is accompanied by acoustic emissions in the frequency band 1 to 100 kH, mostly at boundaries between rock salt and clay or anhydrite [*Spies and Eisenblätter*, 2001]. The frequency of occurrence ranges from ~10 events/hr for small events to short bursts of as many as 200 events/hr every several months. A detailed discussion of creep of rocksalt in domes is given by *Carter et al.* [1993].

## 10. Discussion and conclusions

As shown in Table 1, both ice and salt have much more favorable Grüneisen parameters than ocean, which means that the threshold for detection of ultrahigh-energy neutrinos is much lower for detectors in ice or salt compared with ocean. In addition, ambient noise in the ocean is continuous and much higher than in ice or rock salt, except during brief intervals of acoustic emission due to creep or stick-slip in ice and salt.

At frequencies relevant to future searches for hydrodynamic waves emitted in ultrahigh-energy neutrino-induced cascades (~10 to ~60 kHz), both glacial ice and salt domes look promising. As shown in Table 4, due to the small grain size of South Pole ice (~0.2 cm in the top 600 m compared with ~0.8 cm for NaCl in salt domes), ice scatters less strongly than NaCl, whereas ice, due to its hydrogen-bonded structure, absorbs more strongly than NaCl. The last column combines scattering and absorption to get predicted net attenuation length at 30 kHz.

Laboratory measurements of grain boundary scattering in magnesium, which has a hexagonal crystal structure like ice, and of iron and copper, which have cubic structures like NaCl, agree with theory to within a factor 3 throughout the Rayleigh and stochastic regimes [*Merkulov*, 1957a].

*In-situ* measurements of acoustic absorption in glacial ice were made at frequencies 50 – 200 Hz [*Brockamp and Kohnen,* 1965; *Bentley and Kohnen*, 1976]. Although they are at frequencies too low to be directly applicable to thermoelastic waves from cascades, the values agree with theory for those frequencies. No *in-situ* measurement of scattering, as distinct from attenuation, has been made in glacial ice. *In-situ* acoustic scattering data were obtained in the Hockley Salt Mine (Texas) [*Kirby*, 2004] and the Morsleben (Germany) mine [*Manthei and Eisenblätter*, 2005] and plotted in Fig. 7. From the few *in-situ* data for glacial ice (Fig. 6) and salt (Fig. 7), extrapolated to frequencies 10 to 60 kHz, it is clear that South Pole ice has adequately long absorption and scattering lengths to serve as the host medium for an observatory with up to $10^4$ km$^2$ surface x 1 km depth, and that the Morsleben salt mine, of dimensions ~3 km x ~4 km x ~5 km, may also be suitable for a large observatory.

In order to take the next step in evaluating the merits of domes and South Pole ice, it is essential to make detailed *in-situ* measurements of scattering, absorption, and ambient acoustic noise as a function of depth and frequency in South Pole ice and as a function of frequency and grain size in salt domes. In particular, searches should be made and measurements carried out for mines with mean grain size less than that at Morsleben. In ice the largest uncertainty is the magnitude of attenuation in the firn; in salt domes the largest uncertainty is the magnitude of scattering and absorption in clay, liquid inclusions, and anhydrite layers.

**Acknowledgments**. I thank the National Science Foundation for partial support through the IceCube MRE.

Table 1. Conversion of ionization energy into acoustic energy

| Parameter | Ocean | Ice | NaCl |
|---|---|---|---|
| $T$(°C) | 15°C | -51°C | 30°C |
| $\langle v_L \rangle$ [m s$^{-1}$] | 1530 | 3920 | 4560 |
| $\beta$ [m$^3$ m$^{-3}$ K$^{-1}$] | 25.5 × 10$^{-5}$ | 12.5 × 10$^{-5}$ | 11.6 × 10$^{-5}$ |
| $C_p$ [J kg$^{-1}$ K$^{-1}$] | 3900 | 1720 | 839 |
| peak frequency [kHz] | 7.7 | 20 | 42 |
| Grüneisen constant: $\gamma \equiv \langle v_L \rangle^2 \beta / C_p$ | 0.153 | 1.12 | 2.87 |
| maximum practical useful volume | 10$^5$ km$^2$ x 3 km | 10$^4$ km$^2$ x 1 km | 3 x 4 x 5 km$^3$ |

Table 2. Measured parameters for calculations of absorptivity *vs* frequency and temperature

| reference | structure | frequencies | $U$ (eV) | $\tau_o$ [s] | average $\delta_{max}$ | $\lambda_a$ [km], -51°, $f > 100$ Hz | $\lambda_a$ [km], -46°, $f > 100$ Hz |
|---|---|---|---|---|---|---|---|
| Schiller [11] | monocrystal, 2 angles | 800 to 6000 Hz | 0.58 | 3 x 10$^{-16}$ | 0.0094 | 5.7 | 2.9 |
| Kuroiwa [10] | polycrystal | 177 to 790 Hz | 0.57 | 6.9 x 10$^{-16}$ | 0.0087 | 8.6 | 4.5 |
| Oguro et al. [12] | monocrystal, 8 angles | 400 and 2000 Hz | 0.60 | 1.7 x 10$^{-16}$ | 0.0074 | 11.7 | 5.9 |

Table 3. Grain sizes of NaCl in salt domes

| Location | Mean grain size in samples examined |
|---|---|
| Avery Island, LA | ~7.5 mm |
| Bryan Mound, TX | 2 to 40 mm; average 8 mm |
| Big Hill, TX | 3.7 to 60 mm |
| West Hackberry, LA | 6 to 30 mm |
| Moss Bluff, TX | average 11 mm |
| Bayou Choctaw, LA | 10 to 20 mm (at 0 to 728 m) |
| Zuidwending, Netherlands | 25% have $d$ = 1-3 mm; 75% have $d$ = 3-10 mm |
| Hockley Mine, TX | 6 to 8 mm (with 0.1-1 mm anhydrite layers at grain boundaries) |

Table 4. Summary of predictions for Antarctic ice (0.2 cm) and NaCl (0.75 cm)

| Detector array | $\lambda_s$ at 10 kHz | $\lambda_s$ at 30 kHz | $\lambda_a$ at 10 kHz | $\lambda_a$ at 30 kHz | $\lambda_{atten}$ at 30 Hz |
|---|---|---|---|---|---|
| Ice ($d$ = 0.2 cm) | 2000 km | 25 km | 8-12 km | 8-12 km | ~7 km |
| NaCl ($d$ = 0.75 cm) | 120 km | 1.4 km | 3 x 10$^4$ km | 3300 km | 1.4 km |

**Figure Captions**

**Figure 1**. Absorptivity of laboratory ice, of South Pole ice [*Woschnagg et al.,* 2005], and of NaCl [*Palik*, 1985]. South Pole ice is optimally transparent for Cherenkov light; NaCl is transparent at wavelengths from ~0.5 to 5 $\mu$m, in the tail of the Cherenkov distribution.

**Figure 2**. Acoustic absorptivity in sea water at 4ºC and 1 atmosphere (adapted from *Fisher and Simmons* [1977]).

**Figure 3**. Calculated scattering from air bubbles with diameter ranging from 1 mm near the surface to 0.1 mm at a depth of 1000 m in South Pole ice (from *Price* (1993]). Bubble density is 200 cm$^{-3}$.

**Figure 4**. Scattering as function of frequency and grain diameter in South Pole ice [*Price*, 1993].

**Figure 5.** Structures of ice and NaCl: (upper) View of a plane through ice; an acoustic wave moves protons from one bond site to another by motion of four kinds of defects. (lower) In NaCl, there are no protons; instead, acoustic phonons (quantized lattice vibrations) extract energy by colliding inelastically with the Planck distribution of thermal phonons.

**Figure 6**. Absorptivity in polycrystalline ice calculated from laboratory measurements of internal friction by *Kuroiwa* [1964]. See text for discussion of *in-situ* data in glacial ice (solid triangle and solid circle).

**Figure 7**. Calculated scattering coefficient for grain diameters from 0.2 to 2 cm and absorptivity, which is independent of grain diameter, for NaCl. For discussion of data points, see text.

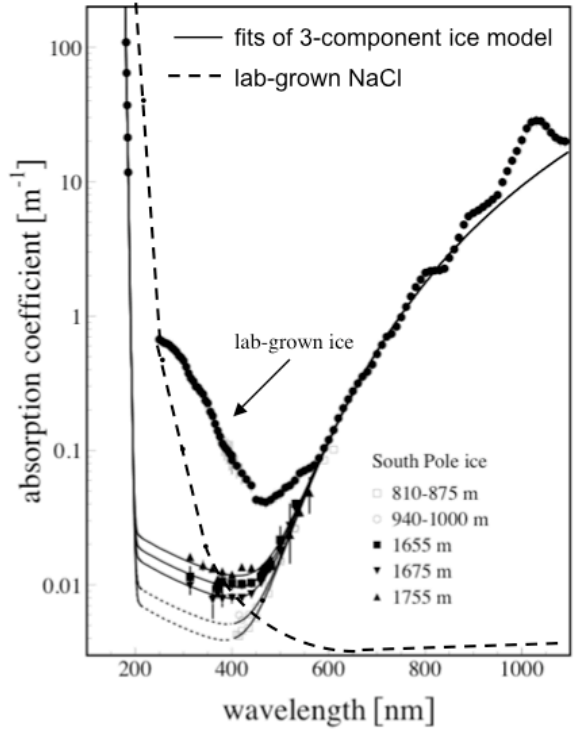

**Figure 1**. Absorptivity of laboratory ice, of South Pole ice at different depths [*Woschnagg et al.*, 2005] and of NaCl [*Palik*, 1985]. South Pole ice is optimally transparent for Cherenkov light; NaCl is transparent at wavelengths from ~0.5 to 5 $\mu$m, in the tail of the Cherenkov distribution.

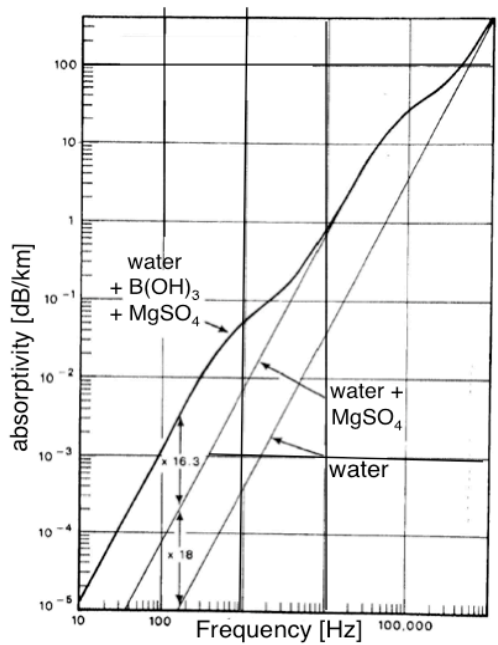

**Figure 2**. Acoustic absorptivity in sea water at 4°C and 1 atmosphere (adapted from *Fisher and Simmons* [1977]).

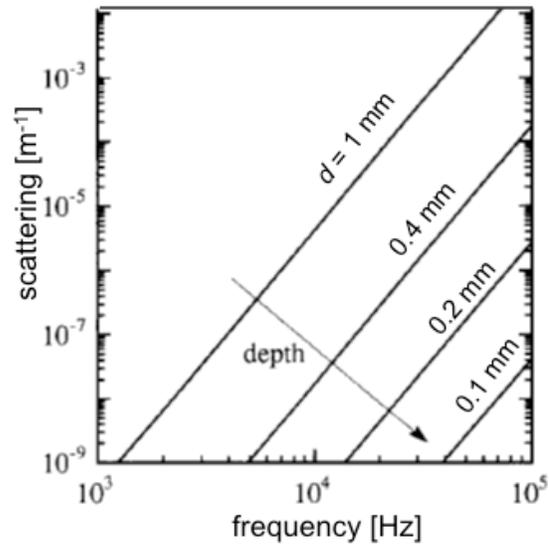

**Figure 3**. Calculated scattering from air bubbles with diameter ranging from 1 mm near the surface to 0.1 mm at a depth of 1000 m in South Pole ice (from *Price* (1993]). Bubble density is 200 cm$^{-3}$.

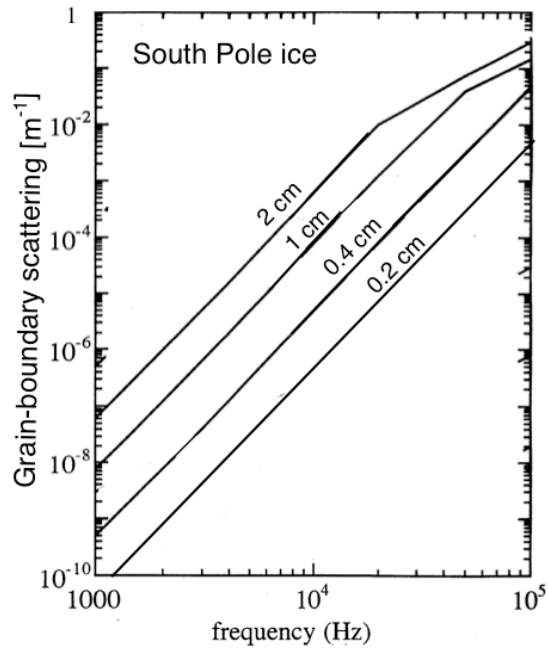

**Figure 4**. Scattering as function of frequency and grain diameter in South Pole ice [*Price*, 1993].

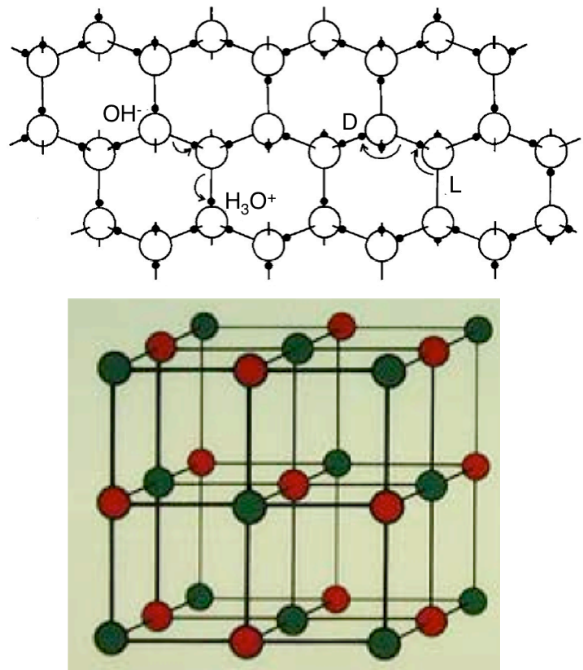

**Figure 5.** Structures of ice and NaCl: (upper) View of a plane through ice; an acoustic wave moves protons from one bond site to another by motion of four kinds of defects. (lower) In NaCl, there are no protons; instead, acoustic phonons (quantized lattice vibrations) extract energy by colliding inelastically with the Planck distribution of thermal phonons.

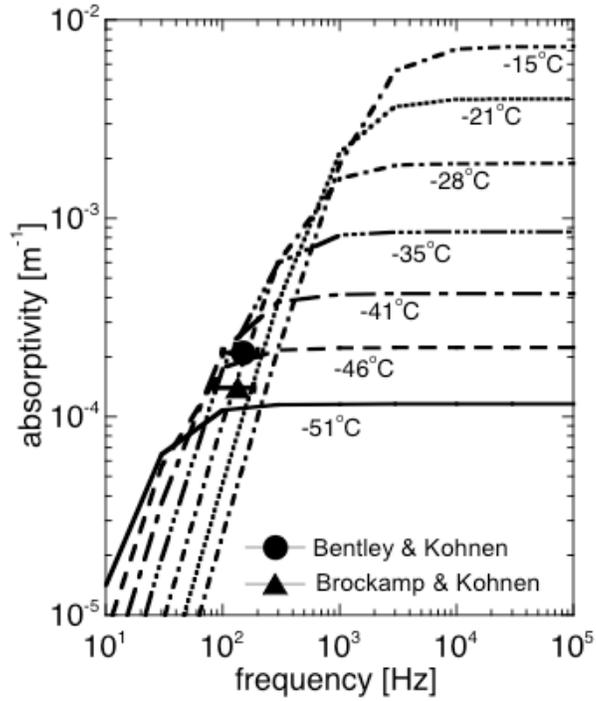

**Figure 6**. Absorptivity in polycrystalline ice calculated from laboratory measurements of internal friction by *Kuroiwa* [1964]. See text for discussion of *in-situ* data in glacial ice (solid triangle and solid circle).

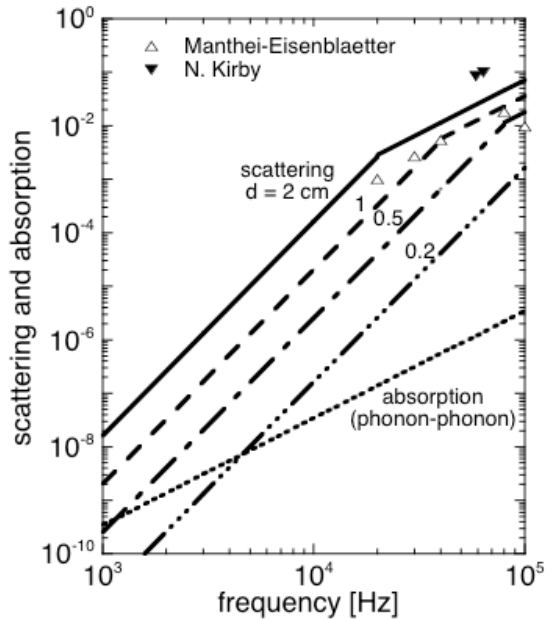

**Figure 7**. Calculated scattering coefficient for grain diameters from 0.2 to 2 cm and absorptivity, which is independent of grain diameter, for NaCl. For discussion of data points, see text.